\documentclass[runningheads]{llncs}
\usepackage[T1]{fontenc}
\usepackage{graphicx}
\usepackage{xcolor}
\usepackage{amssymb}%
\usepackage{amsfonts}
\usepackage{amsmath}
\usepackage{cleveref}
\crefname{lstlisting}{listing}{listings}
\Crefname{lstlisting}{Listing}{Listings}
\usepackage{booktabs}
\usepackage{multirow}
\usepackage{listings}
\usepackage{caption}
\usepackage{array}
\usepackage[htt]{hyphenat}
\usepackage{xspace}
\usepackage{subcaption}
\lstdefinelanguage{Rust}{
  keywords={fn, let, mut, pub, struct, enum, impl, trait, use, match,
  if, else, while, for, in, loop, return, Self},
  morekeywords=[2]{step, stl, println},
  keywordstyle=\color{blue!60!black}\bfseries,
  keywordstyle=[2]\color{red!30!black}\bfseries,
  comment=[l]{//},
  commentstyle=\color{gray},
  stringstyle=\color{black},
  sensitive=true,
}

\newif\ifcomments %

\commentstrue 

\definecolor{burgundy}{rgb}{0.5, 0.0, 0.13}
\definecolor{pink}{rgb}{1.0, 0.0, 1.0}

\usepackage{siunitx}

\newcommand{\round}[1]{%
  \num[scientific-notation=false,round-mode=figures,round-precision=3, group-digits=none]{#1}%
}

\newcommand{\until}[4]{#1  \mathop{\mathcal{U}_{[#3, #4]}}  #2}

\newcommand{\thetool}{\textsc{mstlo}\xspace}

\begin{document}
\title{\thetool:~Efficient Online Monitoring of Signal Temporal Logic}
%
%
\author{Andreas Kaag Thomsen\inst{}\orcidID{0009-0007-0034-7900} \and
  Niels Viggo Stark Madsen\inst{}\orcidID{0009-0000-7383-3466} \and
  Valdemar Tang Evans\inst{}\orcidID{0009-0003-0540-5808} \and
  Thomas Wright\inst{}\orcidID{0000-0001-8035-0884} \and
  Lukas Esterle\inst{}\orcidID{0000-0002-0248-1552}\and
  Peter Gorm Larsen\inst{}\orcidID{0000-0002-4589-1500}
}
\authorrunning{A. K. Thomsen et al.}
%
\institute{Department of Electrical and Computer Engineering, Aarhus
  University, Denmark
  \email{akt@ece.au.dk, nielsviggostark@gmail.com,
    \{valdemar.tang,thomas.wright,lukas.esterle,pgl\}@ece.au.dk}}
\maketitle              
\begin{abstract}
  We present \thetool{} (\textit{mistletoe}), a Rust library for
  high-per\-for\-mance \textbf{\textsc{o}}nline
  \textbf{\textsc{m}}onitoring of \textbf{\textsc{s}}ignal
  \textbf{\textsc{t}}emporal \textbf{\textsc{l}}ogic (STL), with
  Python bindings. The library provides: (i) a unified interface for
  multiple STL semantics, including Robust Satisfaction Intervals
  (RoSI) and Boolean evaluation with early verdicts; (ii) an
  incremental monitoring algorithm based on bottom-up dynamic
  programming with per-operator caching and streaming extremum
  computation for temporal operators; and (iii) an embedded STL
  domain-specific language for both Rust and Python implementations,
  with procedural macros in Rust for static syntax checking.
  Benchmarks show scalability and performance improvements over
  state-of-the-art tools, especially for formulas with large temporal
  depth and deep nesting.
\end{abstract}

\section{Introduction}\label{sec:intro}
The growing deployment of Cyber-Physical Systems (CPSs) motivates the
use of methods for ensuring correctness during
operation~\cite{leeCyberPhysicalSystems2008}. One widely adopted
approach is Runtime Verification (RV), in which a monitor observes
the system's behavior in real time and checks it against formally
specified properties~\cite{leuckerBriefAccountRuntime2009}.
Among the specification languages used in this context, Signal
Temporal Logic (STL)~\cite{malerMonitoringTemporalProperties2004} is
widely used for CPSs, enabling precise specification of timing
constraints, safety requirements, and performance bounds over
real-valued continuous-time signals. STL admits both qualitative and
quantitative semantics, along with offline and online monitoring
algorithms~\cite{deshmukhRobustOnlineMonitoring2017,malerMonitoringTemporalProperties2004}.
In online monitoring, an important goal is to produce a verdict as
early as possible, without waiting for the full temporal depth of the
specification to elapse. This capability is essential for real-time
fault detection but imposes stringent performance requirements: the
monitor must be highly efficient to minimize latency and ensure the
verification process does not lag behind the system itself.

In this paper, we present \textbf{\thetool}, a high-performance RV
library implemented in Rust.
Our tool provides a user-friendly interface for writing STL formulas
and building monitors, while delivering the efficiency and
performance required for production deployment.
\thetool is the first dedicated tool for online monitoring of
multiple STL semantics, which includes delayed verdicts, so-called
Robust Satisfaction Intervals (RoSIs) as well as Boolean early
verdicts. Our tool provides the following key features:
\begin{itemize}
  \item \textbf{Unified Semantics Interface:} We provide a unified
        framework supporting multiple online evaluation semantics via
        efficient stateful streaming algorithms.
  \item \textbf{Embedded Domain-Specific Language (DSL):} We
        introduce a Rust macro-based DSL integrating STL specifications
        directly into source code, with parametrized formulas, syntactic
        sugar, and static syntax checking via procedural
        macros\footnote{\url{https://github.com/rust-lang/rust-analyzer}}. A runtime parser is also provided for the same DSL to allow dynamic formula loading and usage with Python bindings.
  \item \textbf{Efficiency and Benchmarking:} We demonstrate that our
        incremental algorithm is scalable and achieves high throughput
        exceeding state-of-the-art tools, and provide a reproducible
        benchmark suite for hardware-specific performance evaluation.
  \item \textbf{Open-source Rust crate with Python bindings:} Our
        library is available as an open-source Rust crate, ensuring
        accessibility and ease of integration. Additionally, we provide
        Python bindings to facilitate adoption within the widely-used
        Python ecosystem.
\end{itemize}
The source code for \thetool is available at
GitHub\footnote{\url{https://github.com/INTO-CPS-Association/mstlo}},
along with documentation, installation instructions, examples and
comprehensive tests. \thetool has also been successfully used in a
teaching context in the course ``Engineering Digital
Twins''~\cite{fengIncubatorCaseStudy2021} where it has been used to
teach students about how to use RV for monitoring in a practical
setting\footnote{\url{https://github.com/clagms/IncubatorDTCourse}}.
\section{Preliminaries}\label{sec:background}
In the context of RV, we focus specifically on \emph{online
    monitoring}, where the signal is observed incrementally as a stream
of samples rather than as a complete trace available \emph{a priori}.
Additionally, batch processing is also supported, e.g., for post-hoc
analysis of logged data.

\paragraph*{Signal Temporal Logic}
An STL formula $\phi$ is defined recursively by the grammar: \(
\phi ::= \top \mid \mu(x)<c \mid \neg\phi \mid \phi\wedge\psi \mid
\until{\phi}{\psi}{a}{b}
\), where $\mu$ is an atomic predicate over the signal $x$, $\top$
denotes Boolean truth, and $\mathcal{U}_{[a,b]}$ is the time-bounded
\emph{until} operator, requiring $\psi$ to hold at some time within
$[a,b]$ while $\phi$ holds continuously beforehand. Standard derived
operators include disjunction, implication, and the temporal
operators \emph{eventually} ($\lozenge$) and \emph{globally}
($\Box$)\footnote{We restrict our attention to \textit{bounded} temporal operators, i.e., with finite intervals $[a,b]$. Under delayed semantics an unbounded operator never yields a verdict, and under RoSI the interval is trivially $[f_{\inf},f_{\sup}]$ for a large class of STL formulas and thus uninformative~\cite{deshmukhRobustOnlineMonitoring2017}. Supporting them usefully requires a distinct semantics such as nominal RoSI ($\rho_{nom}$)~\cite{deshmukhRobustOnlineMonitoring2017}, which we leave to future work; in practice a sufficiently large $b$ suffices.}.
The \emph{temporal depth}
$H(\phi)$~\cite{yamaguchiRTAMTRuntimeRobustness2024} defines the
maximum future horizon required to evaluate a formula, which dictates
the worst-case buffering requirements for a monitor using delayed semantics.
The expressiveness of STL allows for the compact specification of
complex temporal properties, such as response patterns~\cite{dwyerPatternsPropertySpecifications1999}: ``within the next 100 minutes,
whenever the temperature $T$ exceeds 100 degrees, it must return
below 90 degrees within 5 minutes'' and can be expressed as:
\(
\Box_{[0,100]}\left((T > 100) \rightarrow \lozenge_{[0,5]} (T < 90)\right).
\)
\paragraph*{Supported STL Semantics}
Existing tools often restrict users to a single evaluation strategy.
\thetool, however, implements a unified interface for four distinct
semantic definitions, allowing users to trade off between verdict
expressiveness, verdict latency and performance:

\begin{itemize}
    \item \textbf{Delayed Qualitative:} The standard Boolean
          satisfaction $(s,t) \models
              \phi$~\cite{malerMonitoringTemporalProperties2004}. This requires
          the signal to be fully resolved up to the temporal depth
          $H(\phi)$ before a verdict is emitted.

    \item \textbf{Delayed Quantitative:} The standard robustness
          semantics $\rho(s_t,
              \phi)$~\cite{donzeRobustSatisfactionTemporal2010}, which computes
          a real-valued score indicating the degree of satisfaction or
          violation. Similar to delayed qualitative semantics, this
          requires full signal availability up to $H(\phi)$.

    \item \textbf{Eager Qualitative:} Inspired
          by~\cite{hoOnlineMonitoringMetric2014a}, this semantics leverages
          monotonicity in Boolean and temporal logic to emit early verdicts
          over partial traces via short-circuiting. For example,
          $\Box_{[0,10]} \phi$ yields $\bot$ (false) immediately upon the
          first violation of $\phi$, without waiting for the full interval to elapse.

    \item \textbf{RoSI:} The RoSI semantics $[\rho](\phi,x_
              {[0,i]},t)$~\cite{deshmukhRobustOnlineMonitoring2017} support
          quantitative reasoning over partial traces. Instead of a single
          robustness value $\rho$, the monitor computes an interval
          $[\rho_{\text{min}}, \rho_{\text{max}}]$ enclosing all possible
          future robustness values. A formula is definitively satisfied
          when $\rho_{\text{min}} > 0$ and violated when $\rho_{\text{max}}
              < 0$. The interval is updated incrementally as new signal samples
          arrive and converges to $\rho(s_t, \phi)$ once the full horizon is observed.
\end{itemize}
\section{Efficient Online Monitoring Algorithms}\label{sec:monitoring_algs}
A naive approach to STL monitoring involves re-evaluating the entire
formula from scratch at every time step, a process that scales poorly
with formula complexity and signal length. To address this, \thetool
implements an incremental monitoring algorithm based on a bottom-up
dynamic programming approach, enabling efficient reuse of intermediate
results across timesteps. This approach is well established in RV, and
has previously been applied to e.g. past-time
LTL~\cite{havelundEfficientMonitoringSafety2004} and online
MTL~\cite{hoOnlineMonitoringMetric2014a}, to which we refer
the reader for details.
In our implementation, the specification is internally represented as
an Abstract Syntax Tree (AST), in which each operator maintains its
own cache of intermediate results from its child operators. While this
provides a significant performance boost over the naive method,
temporal operators can still require intensive searches over large
caches to determine a verdict. To mitigate this for \emph{globally}
($\Box$) and \emph{eventually} ($\lozenge$) operators---which are
essentially sliding window minimum and maximum computations---we
incorporate a version of Lemire's
algorithm~\cite{lemireStreamingMaximumminimumFilter2006}.
This optimization significantly reduces both the memory footprint of
the caches and the computational time required to emit a verdict at
each step. It is most effective for non-RoSI semantics; since RoSI
intervals frequently overlap, particularly in nested formulas, the
optimization applies less frequently and still requires a search over
the entire cache, yielding comparatively modest gains.
\section{The \thetool Tool}\label{sec:architecture}
\thetool{} is built around a single generic monitoring core that uses
Rust traits to support multiple semantics with zero-cost static
dispatch. In the following, we describe the key features of the tool.

\subsubsection*{Monitoring Modes}
\thetool supports the four monitoring modes
\texttt{DelayedQuantitative}, \texttt{DelayedQualitative},
\texttt{EagerQualitative}, and \texttt{Rosi} (\Cref{sec:background}).
These modes are implemented via the \texttt{Semantics} configuration,
which determines the type of the output. Additionally,
the monitor can operate in two algorithmic modes: \texttt{Naive},
which re-evaluates the formula at every step, and
\texttt{Incremental} (default), which optimizes performance by only
updating the necessary state (\Cref{sec:monitoring_algs}). The
monitor outputs a verdict for each input timestep.

\subsubsection*{DSL}
\thetool provides an intuitive DSL available both as a compile-time
procedural macro (\texttt{stl!}) for Rust applications and as a
runtime parser. The macro-based approach allows for benefits such as
static syntax checking with expressive error messages; while the
runtime parser supports loading configurations dynamically at runtime
and also enables the DSL for Python bindings.
The DSL supports standard STL syntax as well as syntactic sugar for
common patterns with both shorthand and symbolic notation
(e.g. \texttt{G/globally} and \texttt{\&\&/and}). It also supports
symbolic variables (prefixed with \texttt{\$}) allowing for
parameterized specifications where thresholds can be updated at
runtime. Nested specifications are also supported, enabling modular
and reusable formula definitions.
\begin{lstlisting}[language=Rust, caption={Using the Rust procedural macro DSL to define a formula.}]
let phi1 = stl!(G[0, 5] (temp < $MAX_TEMP));
let phi2 = stl!(pressure > 10.0 -> F[0, 2] valve_open == 1.0);
let phi = stl!(phi1 and phi2);
\end{lstlisting}
\subsubsection*{Building a Monitor}
\thetool utilizes the Builder
pattern~\cite{gammaDesignPatternsElements2011}, allowing users to
configure the formula, semantics, algorithm, variable context, and
synchronization strategy (i.e., how signal data is interpolated when
signals arrive asynchronously) before instantiation. Defaults are
provided for all options.
\begin{lstlisting}[language=Rust, caption={Building a monitor with custom configuration.}]
let vars = Variables::new(); // Define variables context
vars.set("MAX_TEMP", 120.0);
let mut monitor = StlMonitor::builder() // Build the monitor
    .formula(phi)
    .variables(vars)
    .semantics(Rosi)
    .synchronization_strategy(ZeroOrderHold)
    .build()
    .expect("Failed to build monitor");
\end{lstlisting}
\paragraph*{Updating the Monitor}
Once built, the monitor accepts a stream of data points via the
\texttt{update} method, which takes a \texttt{Step} containing the
signal name, value, and timestamp, and returns a
\texttt{MonitorOutput} with any new verdicts, as shown in
\Cref{lst:monitor_update}, which prints \texttt{t=0s:
    RobustnessInterval(-inf, -5.5)} for the given inputs. We adopt a dense-time interpretation, where the monitor can handle irregularly sampled signals.
A verdict can only be produced for a given time step once all
relevant signals have been updated and the monitor has sufficient
information to evaluate the formula according to its semantics.

\begin{lstlisting}[language=Rust, caption={Feeding data and retrieving verdicts.}, label={lst:monitor_update}]
monitor.update(&step!("temp", 125.5, 0s));
monitor.update(&step!("pressure", 15.0, 0s));
let output = monitor.update(&step!("valve_open", 1.0, 0s));
println!("{}", output);
\end{lstlisting}
Batch updates are also supported, allowing users to feed multiple
steps at once and retrieve all resulting verdicts in a single call.
\subsubsection*{Python Bindings}
\thetool includes Python bindings exposed via the \thetool{-python}
package, wrapping the core Rust engine with a Python API and enabling
interactive workflows in environments like Jupyter Notebooks.

\begin{lstlisting}[language=Python, caption={Using the Python interface.}]
formula = mstlo.parse_formula("G[0, 10](x > 5)") # Parse from string
monitor = mstlo.Monitor(formula, semantics="Rosi")
output = monitor.update("x", 6.0, 0.5) # (signal_name, value, timestamp)
print(f"Verdicts: {output}")
\end{lstlisting}
\section{Experimental Results}\label{sec:evaluation}
To evaluate \thetool, we conduct a series of benchmarks. Performance
can vary significantly with the monitored formula (in terms of
nesting and temporal depth), the chosen semantics, and the input
signal (e.g., the signal's frequency and its variation over time, as well as value trends relative to the formula). These benchmarks
provide a representative performance overview showcasing differences
across semantics and central algorithm optimizations.

The input signal is a chirp wave (i.e.\ a sinusoidal signal whose
frequency decreases linearly over time) from $f_0=10^{-1}$\,Hz to
$f_1=10^{-4}$\,Hz, producing values in $[-1, 1]$. We sample the
signal at 1\,Hz for a total of 20,000 samples, although the monitor
is designed to handle variable, unknown sampling rates.
Three formulas are evaluated:
\begin{align*}
    \varphi_1 & = (x < 0.5) \wedge (x > -0.5)
    \\
    \varphi_2 & = \square_{[0,1000]}\bigl(x > 0.5 \to
    \lozenge_{[0,100]}(x < 0.0)\bigr)                    \\
    \varphi_3 & =  \until{(x < 0.5)}{(x < 0.0)}{0}{1000}
\end{align*}
as well as the parameterized temporal formulas
$\square_{[0,b]}\psi_1$, $\lozenge_{[0,b]}\psi_1$, and
$\until{\psi_1}{\psi_2}{0}{b}$ for each $b \in \{1, 100, 200, \ldots,
    5000\}$, where $\psi_1 = (x > 0.0)$ and $\psi_2 = (x < 0.0)$. These
predicates ensure the signal crosses the threshold repeatedly,
exercising both satisfaction and violation, while also preventing
trivial short-circuiting, and ensuring balanced use of Lemire's optimization.
All benchmarks were conducted on an Apple MacBook M4 Pro (24\,GB
RAM), averaging over $M = 50$ runs per formula.
\begin{figure}[!htb]
    \centering
    \includegraphics[alt={Per-sample execution time ($\mu$s) vs.\ temporal upper bound $b$ for $\square_{[0,b]}\psi_1$ and $\until{\psi_1}{\psi_2}{0}{b}$ for \thetool's (Rust) semantics and RTAMT. RoSI and RTAMT's $\mathcal{U}$ are run only for $b\leq 1000$ due to its higher execution time. The graph shows constant scaling for $\square_{[0,b]}\psi_1$ and linear scaling for $\until{\psi_1}{\psi_2}{0}{b}$ in non-RoSI semantics. RoSI shows linear scaling for $\square_{[0,b]}\psi_1$ and quadratic scaling for $\until{\psi_1}{\psi_2}{0}{b}$. RTAMT shows linear scaling for $\square_{[0,b]}\psi_1$ and quadratic scaling for $\until{\psi_1}{\psi_2}{0}{b}$.}, width=0.9\linewidth]{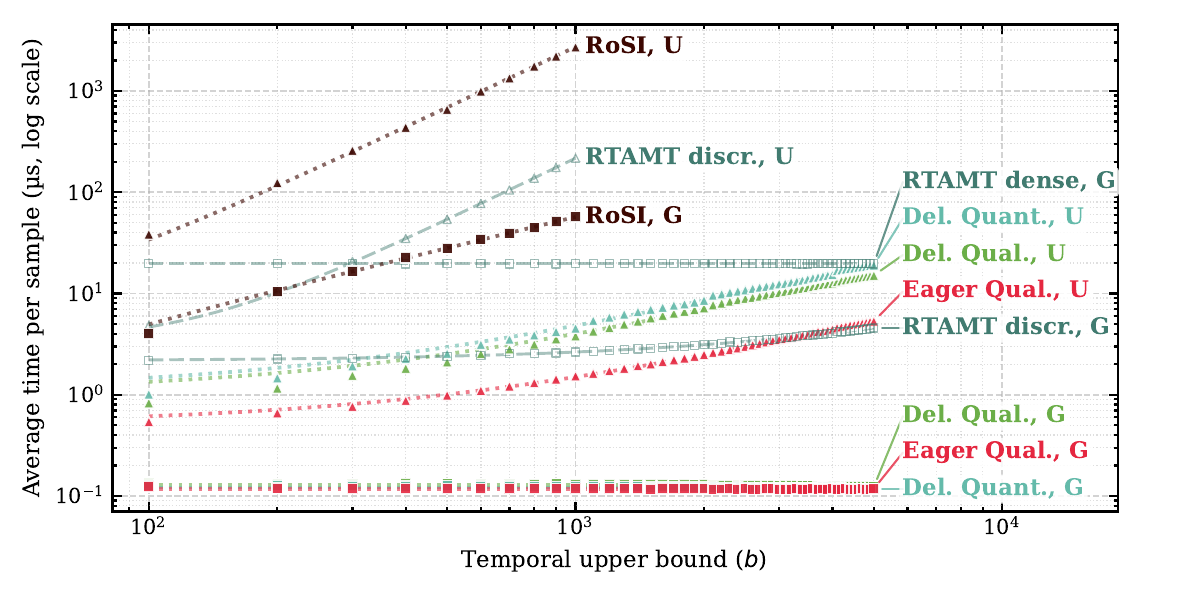}
    \caption{Per-sample execution time ($\mu$s) vs.\ temporal upper bound $b$ for
        $\square_{[0,b]}\psi_1$ and $\until{\psi_1}{\psi_2}{0}{b}$ for
        \thetool's (Rust) semantics and RTAMT. RoSI and RTAMT's $\mathcal{U}$ are run only for $b\leq 1000$ due
        to their higher execution times.}~\label{fig:performance-scaling}
\end{figure}

\paragraph*{Performance of \thetool{}}
\Cref{fig:performance-scaling} shows how execution time scales with
the temporal upper bound $b$. We show only the performance for
$\until{\psi_1}{\psi_2}{0}{b}$ and $\square_{[0,b]}\psi_1$ since
$\lozenge_{[0,b]}\psi_1$ performs similarly to $\square_{[0,b]}\psi_1$.
For delayed and eager semantics, performance is constant for
$\square_{[0,b]}\psi_1$ (due to Lemire's optimization) and linear for
$\until{\psi_1}{\psi_2}{0}{b}$, which does not benefit from it. Eager
qualitative semantics consistently outperforms delayed semantics via
short-circuiting. For RoSI, execution time scales linearly with $b$
for $\square_{[0,b]}\psi_1$ and quadratically for
$\until{\psi_1}{\psi_2}{0}{b}$.

\Cref{tab:comparison} compares the performance across $\varphi_1$,
$\varphi_2$, and $\varphi_3$. The Python bindings introduce only a
small overhead compared to Rust (about $1{-}1.5\times$). The results
agree with the scaling trends observed in
\Cref{fig:performance-scaling}, with execution time increasing with
temporal depth and temporal nesting, and the RoSI semantics being
notably more expensive than the other semantics.
Cache usage reflects formula structure and semantics.
For $\varphi_1$, which has no temporal operators, no caching is needed. For nested formuals ($\varphi_2$), qualitative semantics use minimal caching, while quantitative semantics require substantially more. RoSI incurs the highest cost, but behaves similarly to delayed quantitative semantics when nesting is absent. The cache usage is identical across semantics for $\varphi_3$ since $\mathcal{U}$ does not employ any cache optimization.

\subsection{Case Study: Fault Detection in an Incubator Simulation}
\begin{figure}[!htb]
    \begin{subfigure}[b]{0.45\textwidth}
        \centering
        \includegraphics[alt=Two subplots showing the temperature and heater signals at the top with a lid opening event midway causing a temperature drop,width=\textwidth]{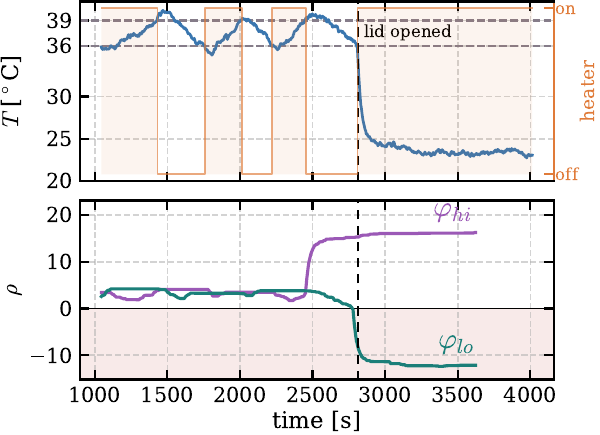}
        \caption{The system signals and robustness values of $\varphi_{lo}$ and $\varphi_{hi}$.}\label{fig:incubator-nominal}
    \end{subfigure}
    \hfill
    \begin{subfigure}[b]{0.45\textwidth}
        \centering
        \includegraphics[alt=The estimated memory usage for each monitor showing a transient startup phase followed by a flattening segment,width=\textwidth]{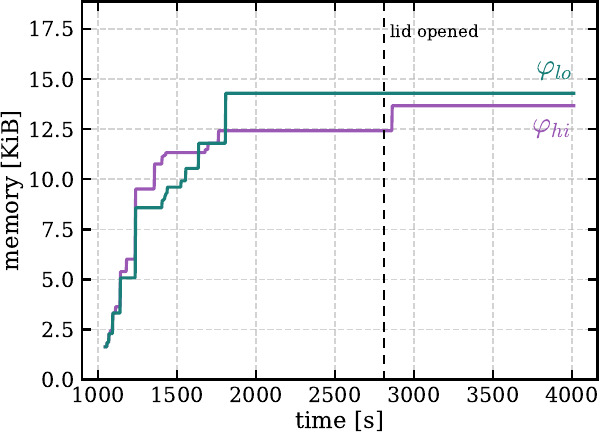}
        \caption{Estimated memory usage of the monitors during operation.}\label{fig:incubator-memory}
    \end{subfigure}
    \caption{Incubator simulation and monitoring results. (a) shows the system signals and the lid opening at $t=2812$s, which causes a temperature drop and the violation of $\varphi_{lo}$. (b) shows that the estimated memory usage of the monitors plateaus after the initial transient phase.}\label{fig:incubator-case-study}
\end{figure}

We now demonstrate \thetool{}'s practical utility through a case study involving an incubator simulation, as used in the course ``Engineering Digital Twins'' at Aarhus University\footnote{\url{https://github.com/clagms/IncubatorDTCourse}}. The incubator is a CPS designed to maintain optimal conditions for the growth of tempeh, by controlling a heating element to keep the temperature within a specified range. Please refer to~\cite{fengIncubatorCaseStudy2021} for a detailed description of the incubator system. In this case study, we monitor the incubator's temperature $T$ to show the system robustness, as specified by the STL formulas
$\varphi_{lo} = \square_{[0,30]}(T \leq \text{\$min\_T} \to \lozenge_{[0,360]}(T \geq \text{\$min\_T}))$ and $\varphi_{hi} = \square_{[0,30]}(T \geq \text{\$max\_T} \to \lozenge_{[0,360]}(T \leq \text{\$max\_T}))$, which are the typical response patterns~\cite{dwyerPatternsPropertySpecifications1999}. A fault is injected into the system by simulating the opening of the box lid, which causes a prolonged temperature drop. In \Cref{fig:incubator-case-study} we show the results of monitoring these formulas as well as the estimated memory usage of the monitor objects. This is an approximation: memory is computed from the reported capacity of the monitor's internal data structures rather than measured at the allocator level, and so can undercount true usage.
In \Cref{fig:incubator-nominal} we observe that the monitor correctly identifies both satisfaction and violation, demonstrating its effectiveness in real-time fault detection within a CPS context. Moreover, \Cref{fig:incubator-memory} shows that memory usage remains bounded throughout monitoring.

%

\newcommand{\delquantOnemu}{\round{0.155997       }}
\newcommand{\delquantOnestd}{\round{0.002518}}
\newcommand{\delquantTwomu}{\round{0.286654       }}
\newcommand{\delquantTwostd}{\round{0.005004}}
\newcommand{\delquantThreemu}{\round{4.390608  }}
\newcommand{\delquantThreestd}{\round{0.051660}}
\newcommand{\delquantHimu}{\round{0.394516}}
\newcommand{\delquantHistd}{\round{0.072551}}
\newcommand{\delquantLomu}{\round{0.329635}}
\newcommand{\delquantLostd}{\round{0.015529}}
\newcommand{\delqualOnemu}{\round{0.159063}}
\newcommand{\delqualOnestd}{\round{0.002281}}
\newcommand{\delqualTwomu}{\round{0.288891}}
\newcommand{\delqualTwostd}{\round{0.003654}}
\newcommand{\delqualThreemu}{\round{3.731765}}
\newcommand{\delqualThreestd}{\round{0.013879}}
\newcommand{\eagerqualOnemu}{\round{0.162903}}
\newcommand{\eagerqualOnestd}{\round{0.003012}}
\newcommand{\eagerqualTwomu}{\round{0.268019}}
\newcommand{\eagerqualTwostd}{\round{0.003344}}
\newcommand{\eagerqualThreemu}{\round{1.123278}}
\newcommand{\eagerqualThreestd}{\round{0.010233}}
\newcommand{\rosiOnemu}{\round{0.171399}}
\newcommand{\rosiOnestd}{\round{0.004250}}
\newcommand{\rosiTwomu}{\round{469.195046}}
\newcommand{\rosiTwostd}{\round{5.050500}}
\newcommand{\rosiThreemu}{\round{2725.987724}}
\newcommand{\rosiThreestd}{\round{6.126930}}
\newcommand{\delquantOnemuPy}{\round{0.24053495799995514}}
\newcommand{\delquantOnestdPy}{\round{0.006821425632750264}}
\newcommand{\delquantTwomuPy}{\round{0.3614899570000034}}
\newcommand{\delquantTwostdPy}{\round{0.010572753851992883}}
\newcommand{\delquantThreemuPy}{\round{4.38294979099976}}
\newcommand{\delquantThreestdPy}{\round{0.07898938516335895}}
\newcommand{\delqualOnemuPy}{\round{0.24125595999987579}}
\newcommand{\delqualOnestdPy}{\round{0.003960989073903319}}
\newcommand{\delqualTwomuPy}{\round{0.370056871999509}}
\newcommand{\delqualTwostdPy}{\round{0.006587699056923935}}
\newcommand{\delqualThreemuPy}{\round{3.751058626999906}}
\newcommand{\delqualThreestdPy}{\round{0.05208707874741228}}
\newcommand{\eagerqualOnemuPy}{\round{0.23536045899913915}}
\newcommand{\eagerqualOnestdPy}{\round{0.004468601374213222}}
\newcommand{\eagerqualTwomuPy}{\round{0.3401842920000035 }}
\newcommand{\eagerqualTwostdPy}{\round{0.005462625882723063}}
\newcommand{\eagerqualThreemuPy}{\round{1.2024186650000956 }}
\newcommand{\eagerqualThreestdPy}{\round{0.012020900352798888}}
\newcommand{\rosiOnemuPy}{\round{0.24463942000056704}}
\newcommand{\rosiOnestdPy}{\round{0.0038850918626920244}}
\newcommand{\rosiTwomuPy}{\round{513.3194592910002}}
\newcommand{\rosiTwostdPy}{\round{5.465934226992081}}
\newcommand{\rosiThreemuPy}{\round{2649.0383794190016}}
\newcommand{\rosiThreestdPy}{\round{81.07495871997298}}
\newcommand{\delquantHimuPy}{\round{0.4266481699801145}}
\newcommand{\delquantHistdPy}{\round{0.01706329092343612}}
\newcommand{\delquantLomuPy}{\round{0.41819175927373864}}
\newcommand{\delquantLostdPy}{\round{0.009628231625693602}}

\newcommand{\rtamtOnemu}{\round{3.095}}
\newcommand{\rtamtOnestd}{\round{0.0576}}
\newcommand{\rtamtTwomu}{\round{4.755}}
\newcommand{\rtamtTwostd}{\round{0.1214}}
\newcommand{\rtamtThreemu}{\round{218.4}}
\newcommand{\rtamtThreestd}{\round{2.418}}
\newcommand{\rtamtHimu}{\round{4.411685176718288}}
\newcommand{\rtamtHistd}{\round{0.4670788267862825}}
\newcommand{\rtamtLomu}{\round{4.2666986853366184}}
\newcommand{\rtamtLostd}{\round{0.09166312273791276}}
\newcommand{\rtamtOnemuDense}{\round{106.43683328686168}}
\newcommand{\rtamtOnestdDense}{\round{0.3452019391314446}}
\newcommand{\rtamtTwomuDense}{\round{77.32853562402308}}
\newcommand{\rtamtTwostdDense}{\round{0.2771380432720254}}
\newcommand{\rtamtThreemuDense}{--}
\newcommand{\rtamtThreestdDense}{--}
\newcommand{\rtamtHimuDense}{\round{37.55184528811332}}
\newcommand{\rtamtHistdDense}{\round{0.972463529778326}}
\newcommand{\rtamtLomuDense}{\round{37.82270260889743}}
\newcommand{\rtamtLostdDense}{\round{0.7968528368952238}}

\begin{table}[!htb]
    \centering
    \caption{Per sample mean and standard deviation ($\mu$s) for \thetool{} across semantics and for RTAMT's discrete-time monitor. $K$ is the number of elements in the caches for all temporal operators in the formula. RTAMT supports only delayed quantitative semantics, and $\mathcal{U}$ is unsupported for dense time.}\label{tab:comparison}
    \setlength{\tabcolsep}{3pt}
    \scriptsize
    \begin{tabular}{c l c >{\color{black!50}}c c >{\color{black!50}}c c c >{\color{black!50}}c c >{\color{black!50}}c}
        \toprule
                                          &                      & \multicolumn{5}{c}{\textbf{\thetool}} & \multicolumn{4}{c}{\textbf{RTAMT (Online)}}  {}                                                                                                                                                                                        \\
        \cmidrule(lr){3-7} \cmidrule(lr){8-11}
                                          &                      & \multicolumn{2}{c}{\textbf{Rust}}     & \multicolumn{2}{c}{\textbf{Python}}             & \textbf{Avg./Max}   & \multicolumn{2}{c}{\textbf{Discr. Time}} & \multicolumn{2}{c}{\textbf{Dense Time}}                                                                             \\
        \cmidrule(lr){3-4} \cmidrule(lr){5-6} \cmidrule(lr){7-7} \cmidrule(lr){8-9} \cmidrule(lr){10-11}
        \textbf{$\boldsymbol{\varphi_i}$} & \textbf{Semantics}   & \textbf{Mean}                         & \textbf{Std.}                                   & \textbf{Mean}       & \textbf{Std.}                            & \textbf{$\mathbf{K}$}                   & \textbf{Mean} & \textbf{Std.}  & \textbf{Mean}      & \textbf{Std.}       \\
        \cmidrule(lr){1-7} \cmidrule(lr){8-11}
        \multirow{4}{*}{$\varphi_1$}
                                          & \textit{Del. Quant.} & \delquantOnemu                        & \delquantOnestd                                 & \delquantOnemuPy    & \delquantOnestdPy                        & 0/0                                     & \rtamtOnemu   & \rtamtOnestd   & \rtamtOnemuDense   & \rtamtOnestdDense   \\
                                          & \textit{Del. Qual.}  & \delqualOnemu                         & \delqualOnestd                                  & \delqualOnemuPy     & \delqualOnestdPy                         & 0/0                                     & --            & --             & --                 & --                  \\
                                          & \textit{Eager Qual.} & \eagerqualOnemu                       & \eagerqualOnestd                                & \eagerqualOnemuPy   & \eagerqualOnestdPy                       & 0/0                                     & --            & --             & --                 & --                  \\
                                          & \textit{RoSI}        & \rosiOnemu                            & \rosiOnestd                                     & \rosiOnemuPy        & \rosiOnestdPy                            & 0/0                                     & --            & --             & --                 & --                  \\
        \cmidrule(lr){1-7} \cmidrule(lr){8-11}
        \multirow{4}{*}{$\varphi_2$}
                                          & \textit{Del. Quant.} & \delquantTwomu                        & \delquantTwostd                                 & \delquantTwomuPy    & \delquantTwostdPy                        & 105/686                                 & \rtamtTwomu   & \rtamtTwostd   & \rtamtTwomuDense   & \rtamtTwostdDense   \\
                                          & \textit{Del. Qual.}  & \delqualTwomu                         & \delqualTwostd                                  & \delqualTwomuPy     & \delqualTwostdPy                         & 3/4                                     & --            & --             & --                 & --                  \\
                                          & \textit{Eager Qual.} & \eagerqualTwomu                       & \eagerqualTwostd                                & \eagerqualTwomuPy   & \eagerqualTwostdPy                       & 3/4                                     & --            & --             & --                 & --                  \\
                                          & \textit{RoSI}        & \rosiTwomu                            & \rosiTwostd                                     & \rosiTwomuPy        & \rosiTwostdPy                            & 1111/1202                               & --            & --             & --                 & --                  \\
        \cmidrule(lr){1-7} \cmidrule(lr){8-11}
        \multirow{4}{*}{$\varphi_3$}
                                          & \textit{Del. Quant.} & \delquantThreemu                      & \delquantThreestd                               & \delquantThreemuPy  & \delquantThreestdPy                      & 1952/2002                               & \rtamtThreemu & \rtamtThreestd & \rtamtThreemuDense & \rtamtThreestdDense \\
                                          & \textit{Del. Qual.}  & \delqualThreemu                       & \delqualThreestd                                & \delqualThreemuPy   & \delqualThreestdPy                       & 1952/2002                               & --            & --             & --                 & --                  \\
                                          & \textit{Eager Qual.} & \eagerqualThreemu                     & \eagerqualThreestd                              & \eagerqualThreemuPy & \eagerqualThreestdPy                     & 1952/2002                               & --            & --             & --                 & --                  \\
                                          & \textit{RoSI}        & \rosiThreemu                          & \rosiThreestd                                   & \rosiThreemuPy      & \rosiThreestdPy                          & 1952/2002                               & --            & --             & --                 & --                  \\
        \cmidrule(lr){1-7} \cmidrule(lr){8-11}
        $\varphi_{lo}$                    & \textit{Del. Quant.} & \delquantLomu                         & \delquantLostd                                  & \delquantLomuPy     & \delquantLostdPy                         & 36/79                                   & \rtamtLomu    & \rtamtLostd    & \rtamtLomuDense    & \rtamtLostdDense    \\
        \cmidrule(lr){1-7} \cmidrule(lr){8-11}
        $\varphi_{hi}$                    & \textit{Del. Quant.} & \delquantHimu                         & \delquantHistd                                  & \delquantHimuPy     & \delquantHistdPy                         & 33/76                                   & \rtamtHimu    & \rtamtHistd    & \rtamtHimuDense    & \rtamtHistdDense    \\
        \bottomrule
    \end{tabular}
\end{table}

\section{Related Work \& Comparison}\label{sec:related-work}
RV has a rich history of ensuring system safety through dynamic
analysis~\cite{bartocciRuntimeVerification2015,havelundDSLsRuntimeVerification}.
Dedicated tools for online STL monitoring are, however, sparse. To the best of our
knowledge, the only existing tool optimized for online STL monitoring is the
Python library \textbf{RTAMT}~\cite{yamaguchiRTAMTRuntimeRobustness2024}, which
employs optimized, incremental update strategies for streaming data, though it
currently only supports delayed quantitative semantics. It supports both discrete-time signals with a C++ backend and dense-time signals, which we compare against in \Cref{tab:comparison}
%

For all formulas, \thetool{}-python is faster than RTAMT (Mann--Whitney U,
two-sided, all $p \ll 0.01$), with per-sample means $10{-}13\times$ lower
(${\sim}50\times$ for $\varphi_3$) than their discrete-time monitor and
$90{-}440\times$ lower than their dense-time monitor (for $\varphi_1,\varphi_2,\varphi_{hi},\varphi_{lo}$).
This gap is not solely attributable to \thetool{}'s algorithmic approach;
RTAMT's heavier reliance on Python compared to \thetool{}'s shallow bindings
likely contributes to the overhead. Also, as shown in \Cref{fig:performance-scaling}, RTAMT's discrete-time monitor does
not employ cache optimization and the temporal operators thus scales linearly
($\lozenge$, $\Box$) or quadratically ($\mathcal{U}$) with their temporal
depths. On the contrary, RTAMT's dense-time online monitor does implement
cache optimization (as noticed by the flat line in \Cref{fig:performance-scaling}), but does not support the $\mathcal{U}$ operator, and is written in pure Python, limiting its performance.

Beyond RTAMT, several tools address related use cases. The C++/MAT\-LAB toolbox
\textbf{Breach}~\cite{donzeBreachToolboxVerification2010} provides the
foundation for STL monitoring and analysis and primarily targets falsification,
model testing, and offline analysis. It includes an online module that computes
RoSI, but this is designed around a fixed evaluation point rather than a
sliding window; at each new time step the monitor re-evaluates over the full
accumulated signal rather than updating incrementally, which limits its
applicability to settings where the time horizon is known and fixed in advance.
Within the Rust ecosystem,
\textbf{Banquo}\footnote{\url{https://github.com/cpslab-asu/banquo}} provides a
STL parser and offline monitoring capabilities.

\thetool bridges these gaps by providing an easy-to-use dedicated
online framework for both Rust and Python that supports multiple
online evaluation semantics, and introduces architectural
advancements tailored for high-performance online monitoring.
Furthermore, \thetool{} provides a DSL with both a procedural macro
for Rust as well as a runtime parser.
\section{Conclusion}\label{sec:conclusion}
In this paper, we presented \thetool, a Rust-based library for
efficient online STL monitoring. By combining multiple evaluation
semantics into a unified framework with an incremental algorithm
leveraging Lemire's sliding-window optimization, implemented in Rust,
\thetool{} advances both the performance and usability of STL
monitoring over existing tools. The embedded DSL adds to this by
providing an intuitive way to specify STL formulas, with benefits
such as static syntax checking. Additionally, Python bindings ensure
accessibility. Directions for future work include extending the
implementation to embedded systems platforms, where memory and
computational constraints impose additional requirements on the
monitoring architecture.

\section*{Acknowledgements}
This work was partially supported by the Grundfos Foundation under
the project DT-CORE and by EU Horizon Europe project RoboSAPIENS
under agreement number 101133807.
%
%
%
%
\bibliographystyle{splncs04}
\bibliography{references}
\end{document}